# A Data-driven Multi-fidelity Physics-informed Learning Framework for Smart Manufacturing: A Composites Processing Case Study

Milad Ramezankhani
Materials and Manufacturing
Research Institute
University of British Columbia
Kelowna, Canada
milad.ramezankhani@ubc.ca

Amir Nazemi
Materials and Manufacturing
Research Institute
University of British Columbia
Kelowna, Canada
amir.nazemi@ubc.ca

Apurva Narayan
Department of Computer Science
University of British Columbia
Kelowna, Canada
apurva.narayan@ubc.ca

Heinz Voggenreiter
Institute of Structures and
Design, German Aerospace
Center (DLR)
Stuttgart, Germany
heinz.voggenreiter@dlr.de

Mehrtash Harandi
Department of Electrical and
Computer Systems Engineering
Monash University
Clayton, Australia
mehrtash.harandi@monash.edu

Rudolf Seethaler
Materials and Manufacturing
Research Institute
University of British Columbia
Kelowna, Canada
rudolf.seethaler@ubc.ca

Abbas S. Milani*
Materials and Manufacturing
Research Institute
University of British Columbia
Kelowna, Canada
abbas.milani@ubc.ca

*Abstract*— **Despite the successful implementations of physics-informed neural networks in different scientific domains, it has been shown that for complex nonlinear systems, achieving an accurate model requires extensive hyperparameter tuning, network architecture design, and costly and exhaustive training processes. To avoid such obstacles and make the training of physics-informed models less precarious, in this paper, a data-driven multi-fidelity physics-informed framework is proposed based on transfer learning principles. The framework incorporates the knowledge from low-fidelity (auxiliary) systems and limited labeled data from target (actual) system to significantly improve the performance of conventional physics-informed models. While minimizing the efforts of designing a complex task-specific network for the problem at hand, the proposed settings guide the physics-informed model towards a fast and efficient convergence to a global optimum. An adaptive weighting method is utilized to further enhance the optimization of the model's composite loss function during the training process. A data-driven strategy is also introduced for maintaining high performance in subdomains with significant divergence between low- and high-fidelity behaviours. The heat transfer of composite materials undergoing a cure cycle is investigated as a case study to demonstrate the proposed framework's superior performance compared to conventional physics-informed models.**

## I. INTRODUCTION

Recent advancements in machine learning and, in particular, deep learning have provided unprecedented opportunities in many complex engineering systems [1]. Specifically in smart manufacturing, deep learning models have been successfully implemented for extracting the underlying complex and nonlinear mapping between the manufacturing settings and the final product properties and quality metrics [2]. Despite being highly flexible and computationally powerful, the performance of deep learning models heavily relies on the availability of large high-fidelity datasets (e.g., experimental measurements, sensory data from the factory floor, etc.) Unfortunately, data scarcity often exists in advanced manufacturing applications as costly and time-consuming manufacturing makes the data collection process a highly prohibitive task. Efforts have been made to address the limitations of deep learning models in small data regimes. Transfer learning (TL) [3] and multi-fidelity learning [4], [5] are the two popular approaches often used to alleviate high-fidelity data limitations by leveraging auxiliary sources of related data/information. In particular, TL aims at learning the task of interest (target) with limited data by transferring the knowledge (e.g., in the form of optimized weights of a neural network) from a related task (source) with abundant data [6]–[8]. In smart manufacturing, for instance, Ramezankhani et al. [9] applied TL in autoclave composites processing to train an accurate neural network for a two-hold cure cycle with limited data by initializing the network with the learned weights from a one-hold cure cycle model. Multi-fidelity's objective, on the other hand, is to learn the *correlation* between the source and target tasks [10]. Multi-fidelity learning can be viewed as a sub-category of TL focusing on scenarios where the knowledge is transferred from abundant *low-fidelity* data, e.g., synthetic data generated by fast simulation models, to scarce *high-fidelity* experimental data. The assumption is that the low-fidelity data contains useful knowledge about the underlying behaviour of high-fidelity data and can be used towards learning an accurate model despite the high-fidelity data limitations. This is especially crucial in advanced manufacturing, e.g., aerospace-grade composites processing,

This study was financially supported by the New Frontiers in Research Fund (NFRF) of Canada – Exploration Program.

for which the trained surrogate model that mimics the system's behaviour requires to exhibit a high accuracy performance with minimal room for error in predictions. The violation of this assumption, e.g., the divergence of low- and high-fidelity data behaviour under certain conditions, can negatively affect the performance of the multi-fidelity framework (i.e., negative transfer [8]). Successful applications of multi-fidelity learning in extracting material's mechanical properties from instrumented indentation [11] and learning the hydraulic conductivity for unsaturated flows [10] demonstrated the effectiveness of incorporating low-fidelity data towards improving the accuracy of high-fidelity models.

Physics-informed neural network (PINN) has recently emerged in many engineering applications [12]–[14]. The PINN model is considered as a faster and more efficient alternative to conventional tools for approximating the solution of partial differential equations (PDE), which is primarily used to describe the behaviour of different phenomena in engineering and science [15]. PINN is capable of learning the underlying behaviour of a system of interest using no or minimal labeled data. In essence, PINN models leverage already-established prior domain knowledge, i.e., governing equations and physical laws, as *inductive biases* to constrain the admissible optimization space of the neural network. Thus, it enables the network to quickly converge to an optimal solution, resulting in an accurate generalization performance [12], [14]. In PINN, such prior knowledge and constraints are incorporated in the form of PDEs. They are imposed in the training process by penalizing the network's loss function using a set of regularizers. This way, it ensures that the neural network satisfies the PDE system of the task at hand.

Training of a PINN model, however, is anything but trivial. Unlike similar conventional tools for solving PDEs, e.g., finite element (FE) and finite volume, for which well-established strategies have been developed to ensure stability and convergence for difficult tasks, PINNs usually requires a lot of trial-and-error iterations in order to obtain an optimal setting for a given problem [16]. In other words, since PINNs rely on neural networks to learn and express the governing equations and the corresponding constraints, it leaves researchers with the formidable task of fine-tuning many hyperparameters. The choice of neural network configuration, weight initialization, activation and loss functions, loss terms weighting strategies and the type of optimizer are only part of the decisions that need to be made to construct a suitable PINN model. Existing guidelines in the literature pertaining to the training of popular deep learning tasks such as computer vision and natural language processing seem to be incompatible with the training of PINN models as they may result in ill-trained, non-optimal models [17]. In addition, although PINN models exhibit promising performance in many simple problems, they tend to fail to learn more complex systems [18], [19]. In particular, it has been shown that introducing soft constraints in the loss function in order to incorporate the PDEs results in a very difficult loss landscape to optimize [15].

To overcome the limitations of PINN while leveraging its capabilities in learning complex engineering systems within small data regimes, this paper proposes a data-driven multi-fidelity physics-informed framework. Unlike conventional multi-fidelity physics-informed models (MFPINNs) [10], the proposed framework utilizes governing laws to train *both* low-fidelity and high-fidelity networks. This further reduces the dependency of the framework on labeled data. Additionally, an adaptive weighting method is implemented that remarkably improves the optimization process of the MFPINN's composite loss function. Finally, a data-driven strategy is introduced to enhance the performance of MFPINN in the subdomains where the low- and high-fidelity behaviours diverge significantly.

## II. METHODOLOGY

### A. Physics-informed neural networks

Typically, engineering systems with a PDE constraint can be formulated as:

$$\mathcal{F}(u(x,t)) = 0, \quad x \in \Omega \subset R^d, \quad t \in [0,T] \quad (1)$$

where $u(x,t)$ is the latent solution (state variable), $\mathcal{F}(\cdot)$ denotes a nonlinear differential operator expressing the PDE, $x$ and $t$ are the space and time with $\Omega$ and $T$ representing the spatial domain and time span. In PINNs, the latent function $u(x,t)$ (e.g., temperature field $T(x,t)$ in a heat transfer problem) is inferred by a feed-forward neural network with unknown parameters $\theta$ representing the weights and biases. An optimal set of parameters can be obtained via an optimization problem, i.e., using gradient descent, to minimize a composite loss function in the form of:

$$\mathcal{L}(\theta) := \mathcal{L}_r(\theta) + \mathcal{L}_{u_b}(\theta) + \mathcal{L}_{u_0}(\theta) \quad (2)$$

Here, $\mathcal{L}_r(\theta)$ denotes the loss term that enforces the governing laws and physics imposed by PDEs. It penalizes the PDE ($\mathcal{F}(u)$) residuals at specified *collocation points* ($x^{pde}, t^{pde}$), often selected randomly. $\mathcal{L}_{u_0}(\theta)$ and $\mathcal{L}_{u_b}(\theta)$ are the losses associated with the initial and boundary conditions, respectively. Similar to $\mathcal{L}_r(\theta)$, initial and boundary points need to be defined to minimize the corresponding losses during the training. A trained neural network $f_\theta(x,t)$ with a near-zero $\mathcal{L}(\theta)$ can represent the solution of the nonlinear PDEs for the task of interest. Mean squared error (MSE) is the common loss function for the PINNs loss terms [12].

### B. Multi-fidelity learning and multi-fidelity PINN

The primary objective in multi-fidelity learning is to learn the relationship between the low- and high-fidelity tasks. To ensure that both linear and nonlinear correlations between the two tasks are taken into account, the following formulation is defined:

$$y_H = f(X, y_L) \quad (3)$$

where $f(\cdot)$ is the unknown function that represents linear/nonlinear correlation between the low- and high-fidelity data and $X$ denotes the input data. Neural networks are good candidates for learning the correlation function $f$ as they often exhibit a robust performance in handling complex nonlinearities. The multi-fidelity learning framework proposed in this study is illustrated in Fig. 1. It comprises of two networks: (1) $NN_L$, which is trained to predict the low-fidelity data $y_L$, given the system's inputs $X$, and (2) $NN_H$, which approximates the high-fidelity data $y_H$ by learning the mapping between $y_L$ and $y_H$. Though this framework can yield improvement in predicting $y_H$, it can be hugely affected by the extent of the present nonlinearities and the availability of labeled low- and high-fidelity data. For instance, if the low-fidelity data is supplied through a computationally expensive FE simulation model, generating a large synthetic dataset for training $NN_L$ might not be feasible. This will lead to a poor prediction of $y_L$ which in turn have a negative impact on the performance of $NN_H$. Even in the presence of abundant low-fidelity data, the lack of sufficient high-fidelity instances might result in poor learning of the correlation function. To address the above, in the proposed framework, the vanilla neural networks, $NN_L$ and $NN_H$ are replaced with their physics-informed variants, $PINN_L$ and $PINN_H$. This results in a multi-fidelity physics-informed neural network (MFPINN) framework. In the proposed MFPINN, the dependency on labeled low- and high-fidelity data is minimized, and the model can be trained by relying on the prior knowledge attained from the governing laws and physics of the problem. The role of the available labeled data, however, will remain vital, as shown in the Results section. To learn the parameters of MFPINN, the following composites loss function should be minimized:

$$\mathcal{L}_{MF}(\theta) := MSE_{y_L} + MSE_{pde_L} + MSE_{y_H} + MSE_{pde_H}, \quad (4)$$

$$MSE_{y_L} = \frac{1}{N_{y_L}} \sum_{i=1}^{N_{y_L}} (|y_L^* - y_L|^2), \quad (5)$$

$$MSE_{y_H} = \frac{1}{N_{y_H}} \sum_{i=1}^{N_{y_H}} (|y_H^* - y_H|^2). \quad (6)$$

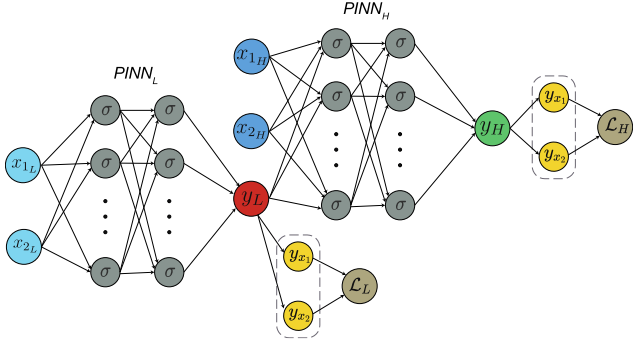

Fig. 1. Schematic of the proposed MFPINN framework. First, $PINN_L$ is trained using the low-fidelity data $(x_L, y_L)$ and PDEs ($\mathcal{L}_L$). Next, to train $PINN_H$, the learned low-fidelity knowledge ($PINN_L$'s predictions) along with the high-fidelity data $(x_H, y_H)$ and PDEs ($\mathcal{L}_H$) are utilized. Automatic differentiation is used to calculate the derivates of the network predictions (yellow circles) needed for calculating the PDE loss terms.

Here, $MSE_{pde_L}$ and $MSE_{pde_H}$ are loss functions defined in (2) and they enforce the physics of the low- and high-fidelity problems. $y_L^*$ and $y_H^*$ denotes the output of $PINN_L$ and $PINN_H$. $MSE_{y_L}$ and $MSE_{y_H}$ take into account the information provided by the labeled data in the training process. It is worth mentioning that $MSE_{pde_L}$ and $MSE_{pde_H}$ loss terms require calculating the derivatives of the networks' approximations (see section II.D). This can be accomplished via automatic differentiation (yellow circles in Fig. 1) for any point within the domain. The networks can be trained simultaneously or in a sequential format with $PINN_L$ being trained first and then be used to predict low-fidelity data for the training of $PINN_H$.

*C. Loss function weighting strategy*

Due to the multi-objective nature of the loss function in PINN models (Equation 2), obtaining the Pareto optima becomes a challenging optimization problem. It has been shown that such settings can lead to stiff gradient flow dynamics which causes unbalanced gradients during the training of PINNs and a poor generalization performance [14]. Additionally, the composite loss function of PINN may result in conflicting gradients [20], which can significantly slow down the convergence of the training and hence increase the number of required iterations. To overcome these issues, in this paper, an adaptive weighting method is employed that uses gradient statistics to update the weights of each term in the PINN loss function [14]. More specifically, at each step of the training, the weight of different loss terms is determined by calculating the corresponding gradient magnitudes $\|\nabla_\theta \mathcal{L}_i(\theta)\|$ and the mean with respect to the network's parameters $\theta$.

*D. Case study: Heat transfer in autoclave composites processing*

Despite the superior mechanical properties, the manufacture of fibre-reinforced polymer composites is a complex multi-step process with a high level of uncertainty. Particularly, in aerospace-grade manufacturing applications, the raw material (typically thermoset prepreg) is cured in an autoclave vessel by applying pre-determined pressure and heat [21]. Maintaining the part's thermal history within the acceptable envelope during the curing process is key to obtaining the desired properties in the cured part. However, due to the complex nature of the curing process, developing a surrogate model using conventional machine learning methods that accurately predicts the thermal behaviour of the part at any given time and location requires a large dataset and an extensive training process. The proposed MFPINN framework offers a data-efficient alternative that learns a robust surrogate model by incorporating the prior knowledge from auxiliary sources of data and the physics of the problem.

The general form of the governing equation for composites heat transfer problem can be written as the following PDE [22]:

$$\frac{\partial}{\partial t}(\rho C_P T) = \frac{\partial}{\partial x}\left(k_{xx}\frac{\partial T}{\partial x}\right) + \frac{\partial}{\partial y}\left(k_{yy}\frac{\partial T}{\partial y}\right) + \frac{\partial}{\partial z}\left(k_{zz}\frac{\partial T}{\partial z}\right) + \dot{Q} \quad (7)$$

where $T$ is the temperature, $\rho$, $Cp$, and $k$ denote the density, specific heat capacity, and the conductivity of the composite part, respectively. $\dot{Q}$ represents the internal heat generation rate in the composite part due to chemical reactions, i.e., polymerization, during the curing cycle. Considering a one-dimensional heat transfer of a *fully-cured* (i.e., no heat generation) homogeneous material and independent physical properties, (7) can be simplified as:

$$\frac{\partial T}{\partial t} - \alpha \frac{\partial^2 T}{\partial x^2} = 0, \quad \alpha = \frac{k}{\rho C_P} \quad (8)$$

where $\alpha$ is the part's thermal diffusivity [23].

In order to model the heat transfer of the cure cycle and achieve a unique solution, two boundary conditions and one initial condition need to be defined. In autoclaves, the part is in contact with the pressurized gas flow (typically nitrogen), which governs the temperature of the part at boundaries and is normally determined by the manufacturer's recommended cure cycle (MRCC) recipe (Fig. 2.a) More specifically, the convective heat transfer between the air and the lower and upper surfaces of the composite part is considered as the boundary conditions of the heat transfer model. In addition, the initial condition is defined as the temperature of the composite part at the initial time step (here 0°C is considered). Thus, the boundary and initial conditions can be stated as:

$$h_b(T|_{x=0} - T_{air}(t)) = k\frac{\partial T}{\partial x}\Big|_{x=0}, \quad (9)$$

$$h_t(T_{air}(t) - T|_{x=L}) = k\frac{\partial T}{\partial x}\Big|_{x=L}, \quad (10)$$

$$T|_{t=0} = T_0, \quad (11)$$

where L is the thickness of the composite part, $T_{air}(t)$ denotes the air temperature at time $t$, $T_0$ is the initial temperature of the part, $h_b$ and $h_t$ are the heat transfer coefficients (HTC) between the air and the bottom and top surfaces of the composite part.

To incorporate the heat transfer PDE constraints into the proposed MFPINN framework, the boundary and initial loss functions ($\mathcal{L}_{u_b}(\theta)$ and $\mathcal{L}_{u_0}(\theta)$) in Equation 2 are defined as:

$$\mathcal{L}_{T_{b1}} = \frac{1}{n}\sum_{i=1}^{n}\left(h_b(T|_{x=0} - T_{air}(t_i)) - k\frac{\partial T}{\partial x}\Big|_{x=0}\right)^2 \quad (12)$$

$$\mathcal{L}_{T_{b2}} = \frac{1}{n}\sum_{i=1}^{n}\left(h_t(T_{air}(t_i) - T|_{x=0}) - k\frac{\partial T}{\partial x}\Big|_{x=l}\right)^2 \quad (13)$$

$$\mathcal{L}_{T_0} = \frac{1}{n}\sum_{i=1}^{n}(T|_{t=0} - T_0(x_i))^2 \quad (14)$$

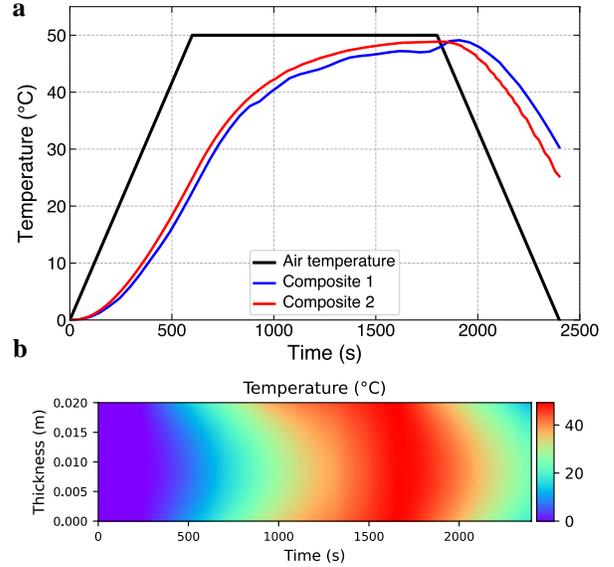

Fig. 2. a) One-hold cure cycle specifications used for the case study. The red and blue curves show the temperature development in the middle section of Composite 1 (low-fidelity model) and Composite 2 (high-fidelity test). b) Temperature distribution of the high-fidelity dataset (Composite 2) during the cure cycle.

The subscript $u$ is replaced by $T$ to represent temperature as the state variable of interest in this case study. In addition, the boundary loss function $\mathcal{L}_{T_b}$ is comprised of two components, namely, $\mathcal{L}_{T_{b1}}$ and $\mathcal{L}_{T_{b2}}$, which impose the boundary constraints on the bottom and top sides of the composite part governed by the autoclave's airflow.

In this case study, two carbon fiber epoxy systems with different physical and mechanical properties (hence, different thermal behaviours) are investigated and their properties are summarized in Table I. Instead of feeding MFPINN with low and high-fidelity data, here, the data from two different composite systems are used with the goal of discovering the correlation among them. The idea is that such correlation between the two materials exists (though in different nature from that of low- and high-fidelity systems) and MFPINN can learn that relationship and utilize it for learning the surrogate model for the material of interest. This has been successfully implemented in TL (i.e., learning from one material to train a surrogate model for another) [6], [9] and, here, the effectiveness of MFPINN in finding such correlations is evaluated. For the sake of generality, we continue to use "low-fidelity" and "high-fidelity" to address the two composite systems. It is assumed that the low-fidelity system (*Composite 1*) contains abundant data, e.g., historical manufacturing data, whereas only a handful of measurements are available for the material of interest (*Composite 2*). Fig. 2.a shows the temperature profile of the two composite parts at their center. An identical one-hold cure cycle is used for both composite parts (black curve).

TABLE I. SUMMARY OF SPECIFICATIONS AND PARAMETERS USED FOR THE CASE STUDY

| Parameter | Value | |
|---|---|---|
| | Composite 1 (Low-fidelity) | Composite 2 (High-fidelity) |
| Part length(mm) | 20 | 20 |
| Density (kg/m³) | 1573 | 1581.26 |
| Conductivity (w/mK) | 0.47 | 0.702 |
| Specific heat capacity (J/kgK) | 967 | 1080.22 |
| HTC - top (W/m²K) | 100 | 100 |
| HTC - bottom (W/m²K) | 50 | 50 |
| Initial temperature (°C) | 0 | 0 |

ABAQUS commercial software was implemented to simulate the heat transfer of the composite parts and solve the corresponding heat transfer PDEs. After mesh consistency and time step stability checks, the number of elements on the composite part and the time discretization were determined as 40 and 0.0015 s, respectively. The maximum allowable temperature change per time step is also selected as 1°C. For training the MFPINN model, all networks have 5 hidden layers equipped with 30 neurons and hyperbolic tangent activation function. ADAM optimizer with a learning rate of 0.001 is used. The learning rate is reduced by a factor of 0.5 once no improvement is observed for 20 epochs. A batch size of 64 and total epochs of 200 are implemented. The adaptive weighting method described in section II.C is utilized for all training processes. A test dataset of 5658 points is used to evaluate the models' generalization performance. All models are constructed and trained in Python using the TensorFlow library.

### III. RESULTS

#### A. Effect of labeled data on PINNs performance

In this section, the effect of incorporating labeled data in the training of a PINN model is investigated. For this case study, the data and specifications from *Composite 2* system (high-fidelity) are employed. 1600 collocation points are uniformly selected in the domain to measure the PDE loss. 20 and 80 initial and boundary points are also chosen to calculate the corresponding losses. 50 labeled data points randomly generated from FE simulation. Fig. 3.a shows the PINN model's absolute error when trained only on PDE. The model exhibits a poor prediction performance around the time 500 s which corresponds to a sharp shift in the boundary condition, i.e., a transition from heating ramp to the hold step (Fig. 2.a). The performance of the PINN model when labeled data is included in the training is shown in Fig. 3.b. The presence of labeled data resulted in a noticeable improvement in the performance of the PINN model, especially near the error-prone regions. Table II summarizes the effect of the labeled dataset size on the PINN generalization accuracy. The relative $L_2$ error is calculated for each case using the following equation:

$$E = \sqrt{\frac{\Sigma_j(y_j^* - y_j)^2}{\Sigma y_j^2}}, \quad (15)$$

where *j* is the index of the test data point.

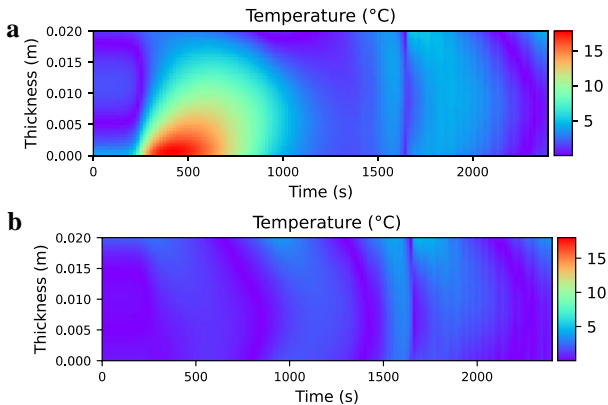

Fig. 3. Effect of labeled high-fidelity data in the PINN model performance. Absolute error between the true temperature and the prediction of the trained PINN model a) with no labeled data (relative $L_2 = 0.13$), and b) with 100 FE labeled data are shown (relative $L_2 = 0.051$). For a fair visual comparison, the same color bar scale is used for both figures.

TABLE II. RELATIVE $L_2$ ($\times 10^{-2}$) ERROR OF PINN MODEL FOR DIFFERENT SIZES OF LABELED DATASET.

| Labeled dataset size | Relative $L_2$ ($\times 10^{-2}$) |
|---|---|
| 10 | 10.03 |
| 50 | 5.1 |
| 100 | 3.16 |
| 200 | 1.89 |
| 400 | 1.67 |

#### B. Multi-fidelity learning

Here, the effectiveness of the multi-fidelity learning approach is investigated. For all the subsequent analyses, the low-fidelity portion of MFPINN framework ($PINN_L$) is trained using the data and governing equations of *Composite 1* system. Specifically, 200 labeled data is utilized to mimic the data abundancy in the source (i.e., low-fidelity) model. Next, in order to evaluate the effect of incorporating the knowledge from the low-fidelity system, the MFPINN model is trained with no labeled high-fidelity data. This leaves the $PINN_H$ model to be trained using the information from the low-fidelity predictions $y_L$ provided by $PINN_L$ and the governing laws incorporated via PDE losses. Fig. 4.b shows the $PINN_H$'s prediction error of *Composite 2*'s temperature distribution. Compared to the case in which no low-fidelity data/PDE is involved (vanilla PINN in Fig. 4.a), the model's performance improved significantly (maximum error reduced to less than 6 °C). The model was able to successfully utilize the low-fidelity knowledge to correct its prediction around the time 500 s where previously a 15°C error was observed. However, with the addition of low-fidelity knowledge, the model's performance seems to decline towards the end of the time domain (2000 s onward). This behaviour can be explained by looking at the low- and high-fidelity temperature developments over time as shown in Fig. 2.a. Though the two curves exhibit relatively similar trends throughout the cure cycle, a noticeable divergence happens towards the end of the cycle, during the *cooldown phase*. This exactly overlaps with the region where MFPINN shows erroneous performance. As described in the next section, this problem will be addressed by incorporating high-fidelity labeled data sampled from the cooldown phase in the training process.

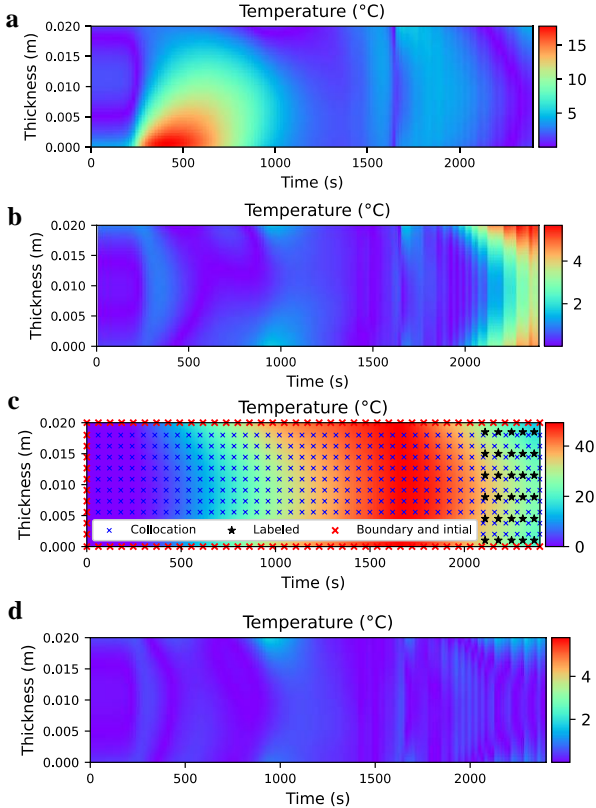

Fig. 4. Performance of MFPINN model in composites' heat transfer case study. Absolute error between the true (FE) temperature and model's prediction is illustrated for a) vanilla PINN, b) MFPINN with no labeled high-fidelity data and d) MFPINN with labeled high-fidelity data. c) shows the experimental design described in section III.C.

### C. Data-driven multi-fidelity learning

A known problem in the training of PINN is its inaccuracies in the vicinity of discontinuities and sharp shifts in the boundary conditions (as observed in section III.A) [13], [24]. One remedy is to locate such areas in the domain, i.e., regions with high training errors, and increase the density of collocation points for further training. Inspired by this idea, a data-driven strategy is introduced here to address the inaccuracies in MFPINN's predictions during the cool-down phase due to the divergence of low- and high-fidelity behaviours. A high-density cloud of labeled high-fidelity points (here, 30 points) is added near the cool-down phase (Fig. 4.c). The idea is that the model can leverage the new information from the labeled high-fidelity data to correct its erroneous predictions arising from the large deviation between low- and high-fidelity data. Fig. 4.d demonstrates the model's new error field. It clearly shows that employing high-fidelity data can effectively bridge the knowledge gap between the low- and high-fidelity data where the deviation is significant. It is worth mentioning that such error-prone regions can be detected and addressed prior to model training by studying the behaviour of low- and high-fidelity systems. A cloud of high-fidelity points can then be used in those subdomains. Table III summarizes the generalization performance of all 4 models investigated in this study. MFPIIN models outperformed conventional PINN models and the dominant performance was achieved by introducing labeled high-fidelity data to the MFPINN model. Fig. 5 also demonstrates the models' prediction of the part's temperature at the center ($L = 1$ cm). The zoom-in view illustrates how the addition of labelled data can prevent the model from overshooting (green and red curves.)

TABLE III. GENERALIZATION PERFORMANCE (RELATIVE $L^2$ ERROR) OF DEVELOPED MODELS.

| Model | Relative $L_2$ ($\times 10^{-2}$) error |
|---|---|
| PINN | 12.9 |
| PINN with labeled data | 5.1 |
| MFPINN | 2.8 |
| **MFPINN with labeled data** | **1.7** |

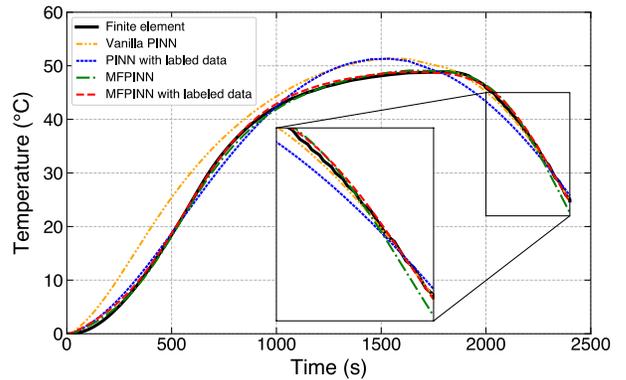

Fig. 5. Prediction of PINN and MFPINN models for part's temperature at the center (Composite 2). FE result is also presented for comparison. The zoom-in view emphasizes the MFPINN model improvement as a result of adding labeled high-fidelity data.

### IV. CONCLUSION AND FUTURE WORK

In this paper, a new multi-fidelity learning framework comprised of two PINN models was proposed. The first network is responsible for learning the underlying behaviour of the low-fidelity data by incorporating both labeled data and governing physical laws. The second network then uses the first network's predictions as an auxiliary source of information towards learning the correlation between the low- and high-fidelity data. High-fidelity governing laws and labeled data are also imposed to guide the model to efficiently converge to a global optimum, resulting in a robust predictive model for the high-fidelity system (task of interest) despite the limited available data. An adaptive weighting method is implemented for the PINN models' composite loss terms to address the issues of stiff gradient flow dynamics and conflicting gradients during the training. The presented results clearly demonstrated that the proposed framework successfully overcomes the shortfalls of vanilla PINN and MPINN in dealing with complex problems in engineering applications.

ACKNOWLEDGMENT

The authors would like to thank the support and helpful comments from the colleagues at the Composites Research Network (CRN) and the University of British Columbia. This study was financially supported by the New Frontiers in Research Fund (NFRF) of Canada – Exploration Program.